\documentclass[11pt]{article}
\usepackage{aaspp4}

\makeatletter

\let\@internalcite\cite
\def\cite{\def\citename##1{##1}\@internalcite}
\def\shortcite{\def\citename##1{}\@internalcite}

\def\reserved@a{LaTeX2e}
\ifx\reserved@a\fmtname
  \usepackage{epsfig}
  \newcommand{\myepsf}[3]%
{\begin{center}\epsfig{file=#2,height=#1,angle=#3}\end{center}}
\else
  \newcommand{\myepsf}[3]%
{\begin{center}\epsfysize=#1 \leavevmode\epsfbox{#2}\end{center}}
\fi

\makeatother

\newcommand{\pderiv}[2]{\frac{\partial #1}{\partial #2}}

\newcommand{\einternal}{e_{\rm int\/}}

\newcommand{\erg}{{\rm\, ergs}}
\newcommand{\inflow}{{\rm in}}
\newcommand{\ism}{{\rm ISM}}
\newcommand{\kel}{{\rm \, K}}
\newcommand{\kmps}{{\rm \: km \, s}^{-1}}
\newcommand{\msolar}{M_\odot}
\newcommand{\nh}{n_{\rm H}}
\newcommand{\nhone}{n_{{\rm H},1}}
\newcommand{\parsec}{{\rm \, pc}}
\newcommand{\percc}{\; {\rm cm}^{-3}}
\newcommand{\yr}{{\rm \, yr}}

\newcommand{\cool}{{\rm cool}}
\newcommand{\steady}{{\rm ss}}
\newcommand{\meanvs}{\left< v_s \right>}

\lefthead{Kimoto \& Chernoff}
\righthead{Supernova Blast Waves: Radiative Instabilities}

\begin{document}

\title{Radiative Instabilities in Simulations of \\ 
       Spherically Symmetric Supernova Blast Waves}

\author{Paul A. Kimoto}
\affil{Department of Physics, Cornell University, Ithaca, NY 14853}

\and 

\author{David F. Chernoff}
\affil{Department of Astronomy, Cornell University, Ithaca, NY 14853}

\begin{center}
accepted by The Astrophysical Journal \\
\copyright 1997 The American Astronomical Society
\end{center}



\begin{abstract}

High-resolution simulations of the cooling regions of spherically symmetric 
supernova remnants
demonstrate a strong radiative instability.  This instability,
whose presence is dependent on
the shock velocity, causes large-amplitude fluctuations in the shock
velocity.  The fluctuations begin almost immediately after the radiative
phase begins (upon shell formation) if the shock velocity lies in the
unstable range; they last until the shock slows to speeds less than
approximately~$130 \kmps$.  We find that shock-velocity fluctuations from
the reverberations of waves within the remnant are small compared to those
due to the instability.  Further, we find (in plane-parallel simulations)
that advected inhomogeneities from the external medium do not interfere
with the qualitative nature of the instability-driven fluctuations.
Large-amplitude inhomogeneities may alter the phases of shock-velocity
fluctuations, but do not substantially reduce their amplitudes.

\end{abstract}




\section{Introduction}   \label{intro}

The simplest description of a supernova remnant involves a spherically
symmetric shell whose expansion is smoothly decelerated by the
surrounding interstellar medium.
Actual remnants appear irregular, and this description must be
supplemented by considering a variety of physical effects
(\cite{chevalier}; \cite{draine-mckee}).  Among the instabilities that have
been investigated are convective motions, the rippling of thin
dense shells, and cooling instabilities affecting radiative shocks.

Much work has been done to investigate instabilities that
may arise during
the initial interactions between a supernova and its
surrounding medium, during which the supernova first begins to decelerate.
Included are convective Rayleigh-Taylor-type instabilities in hollowed-out
Sedov-Taylor solutions (\cite{goodman}) and at the interface between shocked
supernova gas and shocked gas from the surroundings (\cite{cbe};
\cite{chevalier-blondin}), and the effects of expansion into material with
exponentially decreasing density (\cite{luo-chevalier}).  These instabilities
do not prevent the blast wave from entering an adiabatic phase roughly
described by the standard Sedov-Taylor solution.

The dense shell that forms much later, when a remnant enters the radiative
phase (in which the cooling timescale becomes shorter than the remnant age)
is subject to a rippling instability (\cite{vishniac};
\cite{bertschinger}).  When it occurs, the deviations from sphericity
oscillate overstably, and the shell becomes wrinkled.  Linear models
indicate that the oscillations grow as a power law in time.

This paper focuses on the effects of the global cooling overstability
on the evolution of a spherically symmetric supernova remnant.  In this
instability the cooling lengths of radiative shocks oscillate
(\cite{lcs}; \cite{chevalier-imamura}; \cite{imamura-wolff-durisen}).
The phenomenon is essentially one dimensional, in the direction of shock
propagation, and so is easily amenable to numerical investigation.  The
instability develops exponentially rapidly, and (in the context of a
decelerating shock) requires no other external perturbations.
Bertschinger~(\shortcite{bertschinger}) suggested that this cooling
instability could provide the initial seed perturbations to drive the 
slower-growing, later-phase rippling instability.

According to linear perturbative analysis with power-law cooling
laws~${\cal L} \propto \rho^2 T^\alpha$, instability occurs when~$\alpha$
is small,
$\alpha \la 0.8$ (\cite{chevalier-imamura}).
With such cooling laws, fast shocks support long cooling lengths, and slow
shocks require short cooling lengths.  Deviations from steady-state flow
tend to drive the system away from that steady state.

Innes, Giddings,~\& Falle~(\shortcite{innes}) and Gaetz, Edgar,~\&
Chevalier~(\shortcite{gaetz-edgar-chevalier}) studied the evolution of
plane-parallel shocks with nonequilibrium cooling processes.
The simulations, which included the history-dependent effects of
nonequilibrium cooling,
showed that shocks with mean velocities~$v_s \ga 130 \kmps$
are unstable.  
Observations of the Vela supernova remnant
(\cite{raymond-wallenstein-balick}) have found evidence consistent with the
presence of unsteady flow like that produced by this
instability---unusually broad line widths (of \ion{Si}{2} and \ion{Mg}{2})
and a
discrepancy between the shock ram pressure and the postshock thermal
pressure.

Cioffi, McKee,~\& Bertschinger~(\shortcite{cmb}) gave a global picture of the
evolution of a spherically symmetric supernova remnant.  Although this
treatment of the entire remnant interior over most of the life of the remnant
was not concerned with the details of shell formation, its supernova models
provide a context for studying the instability.

In the calculations we discuss below, like 
Cioffi et al.~(\shortcite{cmb}) 
we simulate a~$0.931 \times 10^{51} \erg $ explosion 
into a spherically symmetric interstellar medium.  
We adopt a simple cooling law~${\cal L} = \nh^2 f(T)$, 
where~$\nh$ is the density of hydrogen nuclei 
and $f(T)$~is a piecewise power-law fit 
to the results of Raymond, Cox,~\& Smith~(\shortcite{rcs}), 
altered to turn off at low temperatures (cf.~Appendix~A).  
We assume the gas to be completely ionized, 
and to include helium in a 1:10 ratio to hydrogen.
This cooling law should allow us to capture the features of the cooling
that lead to oscillations of the cooling column.

When the remnant is young, the gas is hot, the cooling rate is small, and so
the remnant is roughly adiabatic.
The remnant changes from adiabatic to radiative near the time at which
the remnant age equals the cooling time of the gas at the shock
front.  When cooling sets in, the gas behind the shock loses pressure support
and so (1)~the shock speed drops suddenly and (2)~a dense shell builds up
behind the shock.  Between the shock and the dense shell lies the cooling
region.

After a brief transition period, the behavior of the shock position~$r_s$
changes from the adiabatic scaling~$r \propto t^{2/5}$ toward the
``pressure-driven snowplow''
scaling~$r \propto t^{2/7}$~(Cioffi et al.~\shortcite{cmb}). 
Since the timescale of the dynamical cooling instability is comparable to
the cooling time (\cite{chevalier-imamura}),
the effects
of the instability appear at the onset of the radiative phase
if the shock velocity
is sufficiently high (i.e., if allowed by the cooling law
at the shock temperature).  
The shock oscillations persist until the shock speed becomes too slow.

Analytic estimates by Cioffi et al.~(\shortcite{cmb}) give
the time of shell formation as
\begin{equation}
  t_{\rm sf} = 3.61 \times 10^4 E_{51}^{3/14}\nhone^{-4/7} \yr ,
\end{equation}
where $E_{51}$~is the initial explosion
energy in units of~$10^{51} \erg$, $\nhone$~is the interstellar density in
units of~$1 \percc$, and we assume that the metallicity is given by solar
abundances.  During shell formation the shock velocity drops strongly until
some postshock gas ends its period of rapid cooling, at which time this gas
reaches its greatest density.  Then the shock begins to move with the dense
shell in the pressure-driven snowplow stage.  The shock radius and velocity
tend toward
\begin{eqnarray}
  \label{rpds}
  R_s & = & 14 \, E_{51}^{2/7} \nhone^{-3/7}
           \left(\frac{4e}{3} \frac{t}{t_{\rm sf}} - \frac{1}{3}\right)^{3/10}
           \parsec , \\
  \label{vpds}
  v_s & = & 413 \, \nhone^{1/7} E_{51}^{1/14} 
           \left(\frac{4e}{3} \frac{t}{t_{\rm sf}} - \frac{1}{3}\right)^{-7/10}
           \kmps
\end{eqnarray}
(Cioffi et al.\ argue that these offset power laws are better
estimates than the familiar power-law relations).

The possibility for oscillations begins at this time.
In our simulations
we find that this stage begins near the time~$1.4 \, t_{\rm sf}$.
In order to have oscillations, the shock velocity must exceed
some cutoff value~$v_c$.  Adopting the approximate expression~(\ref{vpds})
for~$v_s$ during this stage 
(which, however, appears to be an overestimate at such early portions of this
stage), we obtain the requirement
\begin{equation}   \label{minimum_nh}
  \nhone^2 E_{51} \ga 0.39 \left(\frac{v_c}{130 \kmps}\right)^{14}.
\end{equation}

The oscillations may begin immediately upon shell formation 
(as they do in all
of our simulations with oscillations) and last until the shock velocity drops
to~$v_c$ at 
\begin{equation}
  t_c \approx t_{\rm sf} \left[ 0.09 + 1.44
                \left(\frac{v_c}{130 \kmps}\right)^{-10/7}
                \nhone^{10/49} E_{51}^{5/49}\right].
\end{equation}
The oscillation period is approximately~$2\pi t_\cool $, where
$t_\cool $ is given by the cooling rate at the shock temperature
(\cite{chevalier-imamura}).  Since this cooling time drops rapidly as the
shock slows, typically many cycles occur before the oscillations
cease.

\begin{figure}
\myepsf{6.0in}{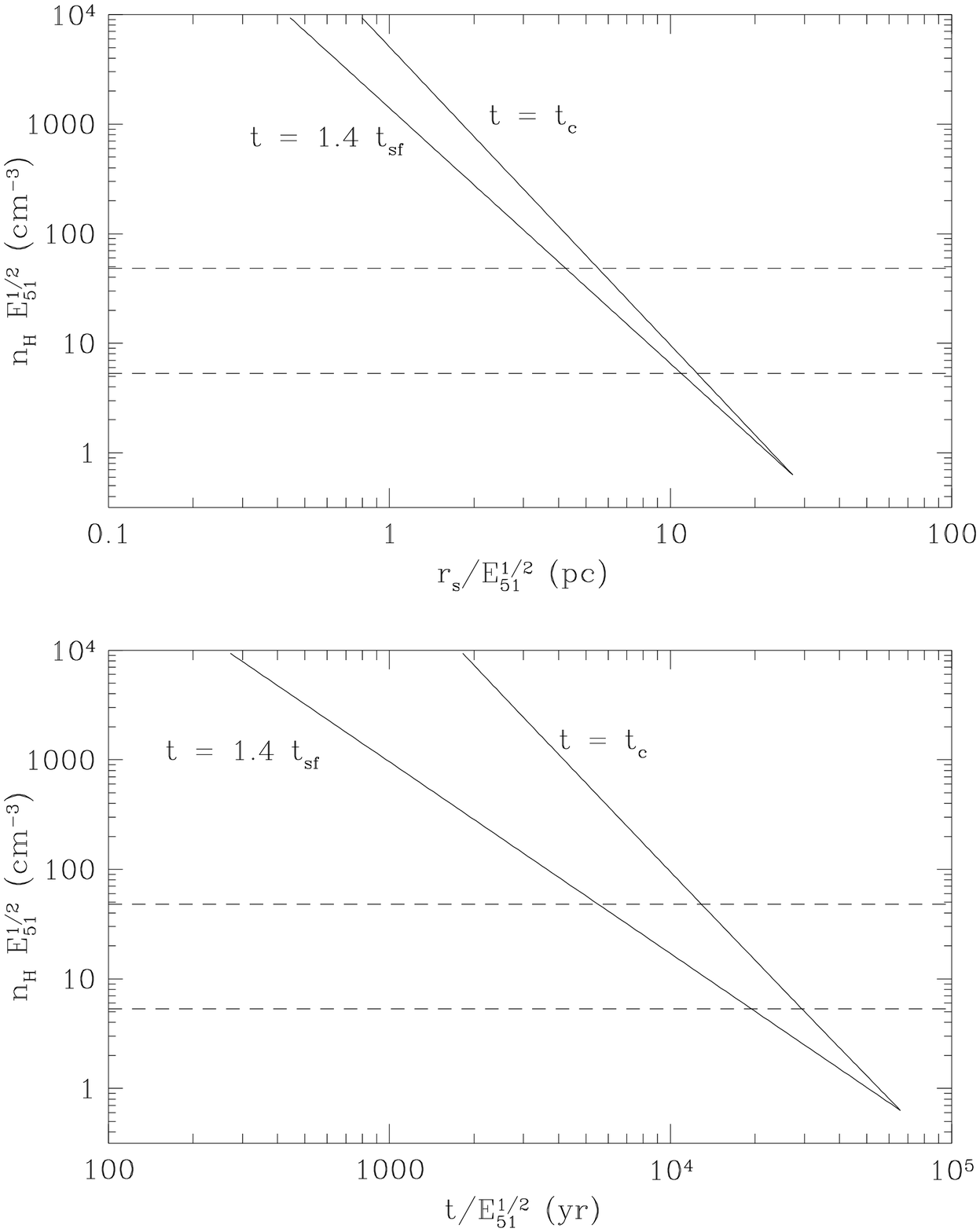}{0}
\caption{The estimated shock radii and time period 
for shock oscillations as functions of interstellar-medium density.
The two solid lines delimit the time interval.  
(Here we assume that the oscillations end when the shock velocity reaches
$130 \kmps$.)  The two horizontal, dotted lines represent the two cases
simulated in which we find oscillations 
($\nh = 5.5 \percc $ and $50 \percc $, both with~$E_{51} = 0.931$).%
\label{parameters}}
\end{figure}

The approximate expression~(\ref{rpds}) for shock radius~$R_s$ during the
pressure-driven snowplow stage gives the corresponding bounds on shock radii
for which oscillations may occur.  The intervals in time and shock radius are
shown in Figure~\ref{parameters},
where the two lines represent the approximate
beginning and end of the intervals.
(We adopt~$v_c = 130 \kmps$ for this simple illustration.)
The need for sufficiently high
shock velocity at the time of shell formation~(\ref{minimum_nh}) is reflected
by the truncation of this region at low values of~$\nh$.

(As noted, the shock velocity is the criterion for determining whether the
shock motion is stable.  With a cooling term of the form~$\rho^2 f(T)$ 
[as we employ], solutions to the equations of motion are invariant under the
transformation~$\rho \rightarrow A\rho$, $p \rightarrow Ap$, $E \rightarrow
E/A^2$, $t \rightarrow t/A$, $r \rightarrow r/A$.  As long as we can neglect
the details of the initial explosion, we can cover the range of values of
explosion energy~$E_{51}$ by varying the interstellar-medium density~$\nh$,
and applying this transformation.
The axes in Figure~\ref{parameters}
are labeled to reflect this convenient result.)

\section{Numerical simulations of shock-velocity oscillations}

All of our spherically symmetric calculations assume an explosion energy
of~$0.931 \times 10^{51} \erg$.  Our inflowing gas has a
temperature~$T_\inflow = 2500 \kel$.  This is of course unrealistic,
but the particular value does not affect the evolution of the shock so
long as the shock remains strong.
The rarest medium we consider has a density~$\nh = 0.1 \percc$.
In the course of the simulations
discussed below, the shock velocity becomes as low as~$49 \kmps$,
with a Mach number of~$6.5$.  
For such a interstellar-medium phase, however, one expects 
the interstellar medium to have~$T_\ism \approx 10000 \kel$.
A~$49 \kmps$ shock expanding into this hotter medium has a modest Mach number
of~$3.3$, but the postshock compression is just 20\% greater than the case of
our $2500 \kel$~medium.
For larger-density interstellar-medium phases one expects ambient
temperatures less than~$2500 \kel$ and smaller discrepancies in the
postshock compression.
In these cases the fact that shocks faster than~$110 \kmps$ heat preshock gas
to temperatures~$T_\inflow \sim 10000\mbox{--}30000 \kel$
(\cite{shull-mckee}) limits the shock strength.  We find that
the maximum deviation in the postshock compression is less
than~$15\%$ from our numerical case.

The cooling law~${\cal L} = \nh^2 f(T)$ uses for the cooling function~$f(T)$
a piecewise power-law fit to the result presented by Raymond et
al.~(\shortcite{rcs}), as discussed above.  To account crudely for the
turnoff of cooling in the recombination zone behind the shock front, we
arrange for the cooling to vanish below a cutoff temperature~$T_c \approx 2
\times 10^4 \kel$.  Our use of this function assumes that the emitting plasma
is in collisional equilibrium.  Because collisional, ionization, and
recombination rates are slow compared to the cooling rate,
the evolution of the ionization state
lags our equilibrium assumption (\cite{innes}; \cite{gaetz-edgar-chevalier}).
At shock speeds greater than~$110 \kmps$, photons emitted by hot postshock
gas preionize the inflowing gas (\cite{shull-mckee}) so that the postshock
gas tends to evolve toward equilibrium, but at lower shock speeds, where
preionization is not so complete, postshock ionization rates are slow and the
cooling may differ greatly from our equilibrium assumption.  
nonequilibrium
simulations~(\cite{innes}; \cite{gaetz-edgar-chevalier}) indicate that
oscillations occur for shock speeds~$\ga 130 \kmps$ and that the
essential feature required is a cooling law with a small local power-law
index~$\alpha$.  Our equilibrium law should capture the correct qualitative
behavior because it has this feature.
With a more realistic cooling law, requiring explicit
evolution of the ionization, simulating the many oscillations shown below
would be a much more computationally taxing task.  Using our simplified
cooling law our longest simulation (the high-density $\nh = 50 \percc$~case)
required approximately six days of computing on a Sun Sparc 10 workstation.

Since we are interested in the dynamics of the cooling region (unlike, for
example, Cioffi et al.~[\shortcite{cmb}]), our numerical
simulations must allow for high resolution between the shock front and the
dense shell.  
The cooling length~$L_\cool$ is quite short compared with the remnant
radius, and as the dense shell expands and slows, the disparity between
these two length scales increases.  Our Eulerian finite-difference method
incorporates a hierarchy of grids that allows high-resolution subgrids to
be placed where needed.  We place high-resolution subgrids in the cooling
region; typically on the order of 100 grid points (evenly spaced, for the
average cooling length) are required to resolve the dynamics that drive the
oscillatory instability.  In cases where the disparity between cooling
length and remnant size is the greatest
(that is, at the lowest speeds for which instability is expected),
we find that it is sufficient to
place such high resolution at the two ends of the cooling region (the shock
front and the transition to the cool, dense shell), which have the steepest
gradients.  (Appendix~A describes the model and numerical methods in more
detail.)  Unfortunately, even at this level of resolution, it is
possible to validate the results of the simulations with rigorous
convergence tests (e.g., \cite{kc}) only for short time intervals.  
Experimentation with simulations with
different resolutions indicates, however, that (1)~we are able to identify
when the instability is and is not present, and (2)~we can readily
determine the relative
amplitudes and periods of oscillation to~$\la 10\%$.

For simplicity we simulate only a portion of the remnant's interior.  
For our first cases we start the simulations at an age
well within the remnant's adiabatic phase,
with the exact Sedov-Taylor blast-wave solution for adiabatic,
point-like explosions~(\cite{sedov}).
For an interior boundary condition
we set the flow variables equal to the values
given by the Sedov-Taylor solution.
We impose this condition at a radius small enough so that it has no
effect on the evolution of the region close to the shock.

\begin{figure}
\myepsf{6.5in}{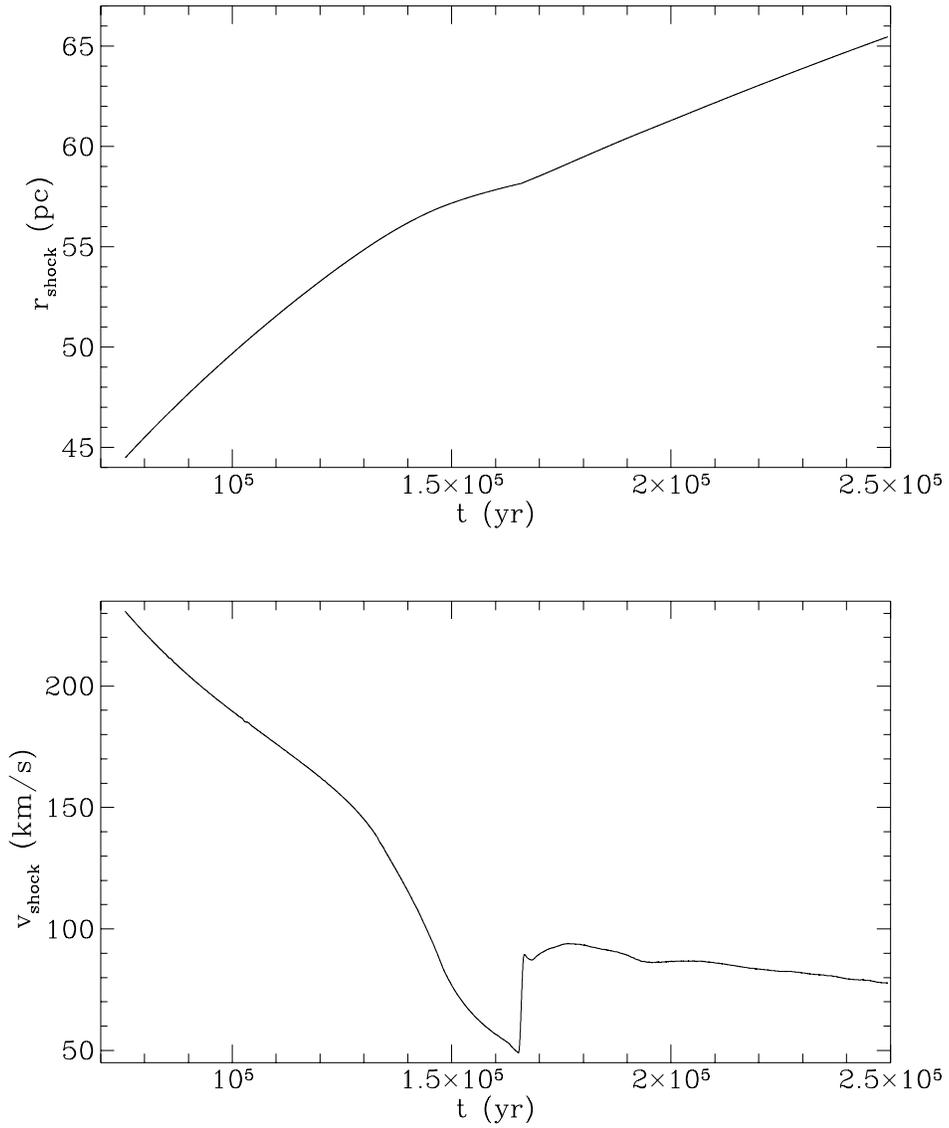}{0}
\caption{Shock position and velocity as a function of remnant age
for an interstellar-medium density~$\nh = 0.1 \percc $.  
\label{case0.1}}
\end{figure}

\begin{figure}
\myepsf{6.5in}{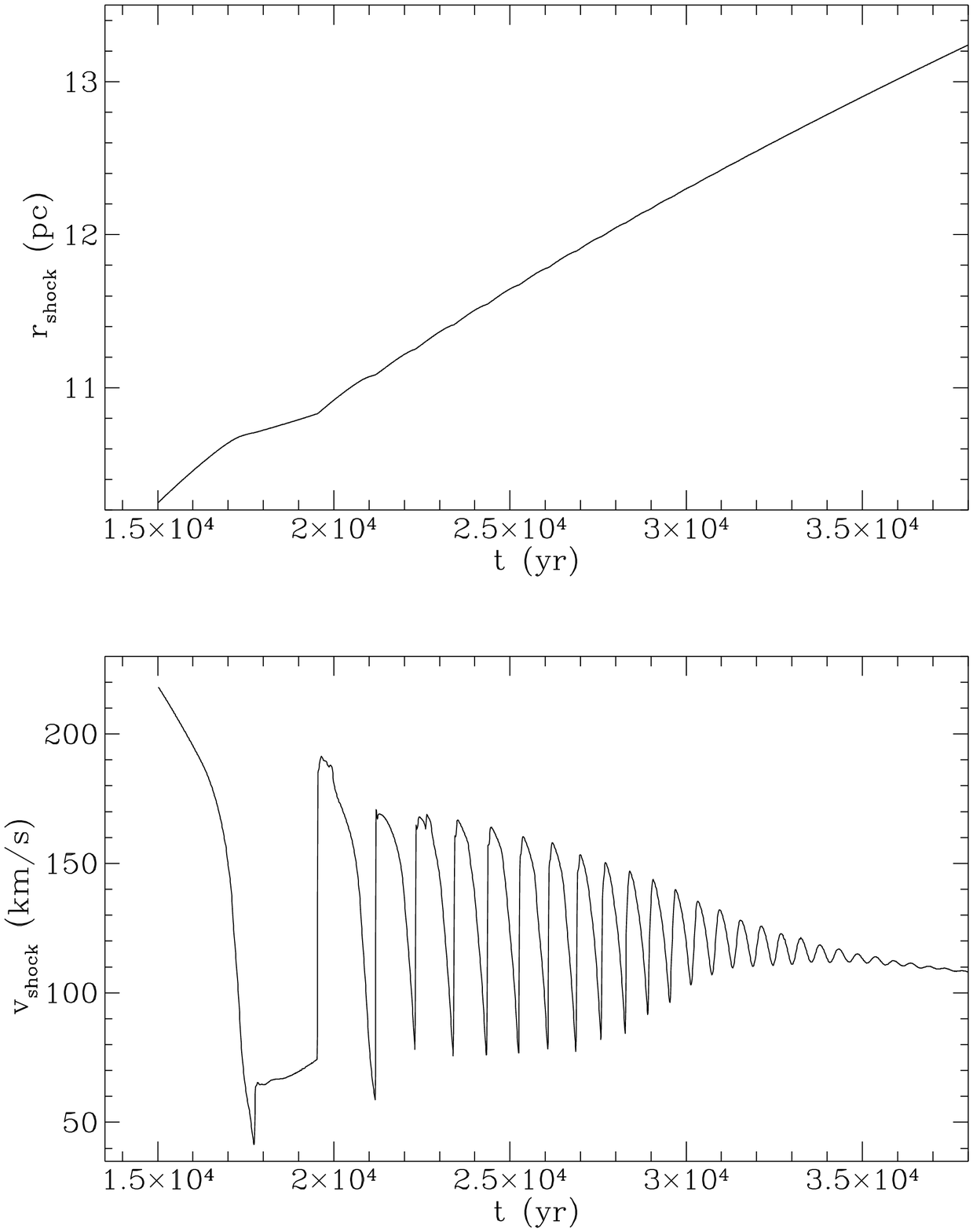}{0}
\caption{Shock position and velocity
for ISM density~$\nh = 5.5 \percc $.
\label{case5.5}}
\end{figure}

\begin{figure}
\myepsf{6.5in}{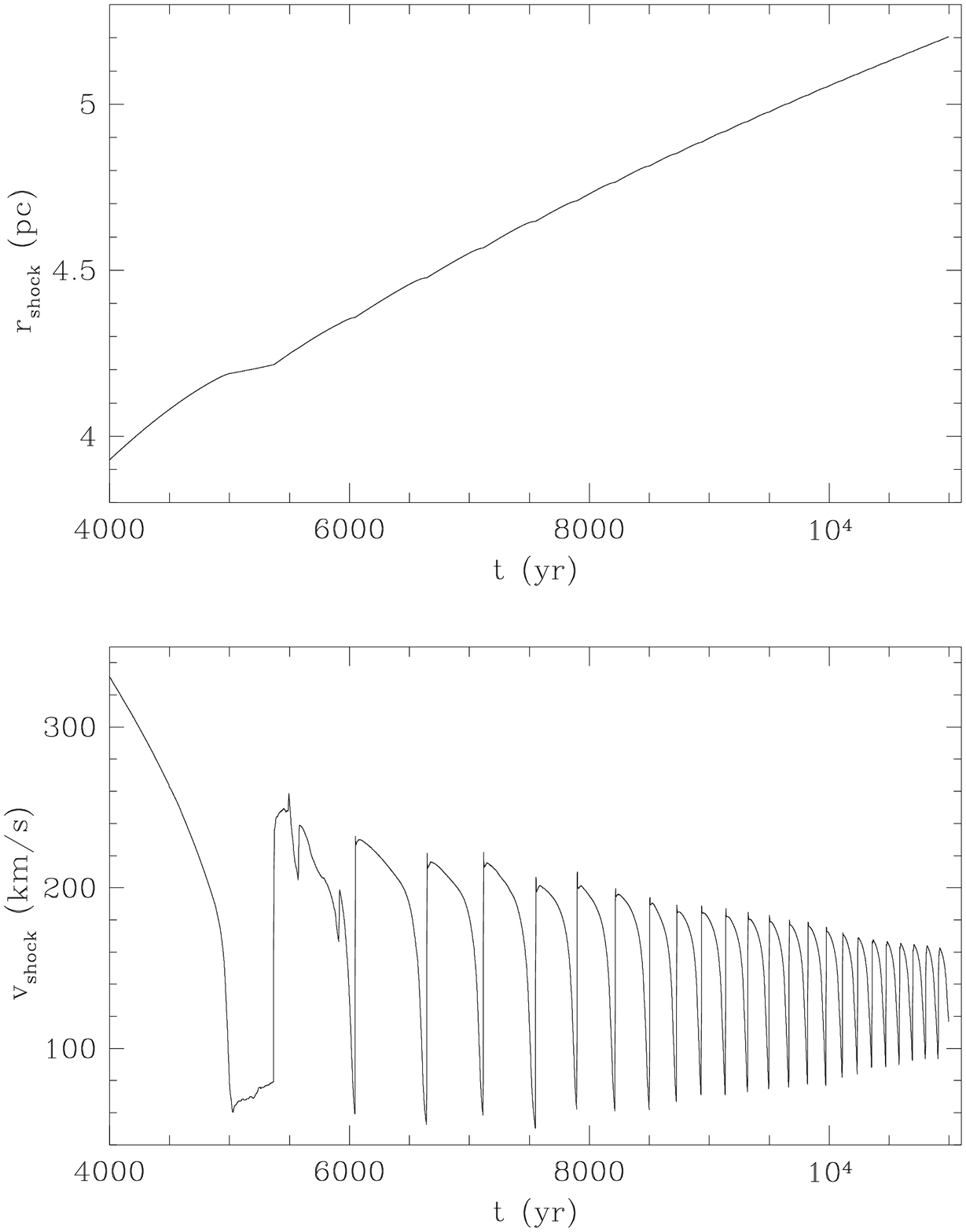}{0}
\caption{Shock position and velocity 
for ISM density~$\nh = 50 \percc $.
\label{case50.0}}
\end{figure}

In Figures~\ref{case0.1}, \ref{case5.5}, and \ref{case50.0}
we show the evolution of the shock position~$r_s$ and
shock speed~$v_s$ for three choices of the interstellar medium density.
The lowest value, $\nh = 0.1 \percc$, is
the value used by Cioffi et al.~(\shortcite{cmb}).  
The other two values, $\nh = 5.5$ and $50 \percc$,
are chosen so that their shock temperatures at the time of shell formation
lie in different power-law regions of our cooling law:
the power-law index changes from~$\alpha \approx -2.2$
to~$\alpha \approx -0.1$ at $T = 5 \times 10^5 \kel$, 
corresponding to a shock velocity of~$190 \kmps$.
For the $\nh = 0.1 \percc$~simulation shown 
we use over the bulk of the remnant 
a grid size~$\Delta r = 0.0064 \parsec$ ($\Delta r/r_s \approx 0.012\%$),
and in the neighborhood of the shock,
a smaller grid size~$\Delta r = 0.0008 \parsec$ (a factor of~$8$ smaller).
The interior computational boundary, at which the Sedov-Taylor solution is
imposed, is at~$r = 31 \parsec$.
For the $\nh = 5.5 \percc$~simulation, 
we use~$\Delta r = 0.0016 \parsec$ ($\Delta r/r_s \approx 0.014\%$) 
over the bulk of the remnant, 
and on the cooling region~$\Delta r$ 
varies from $0.0001 \parsec$ to~$0.00005 \parsec$ as necessary to maintain
$\Delta r/L_\cool \approx 1\%$.  (In addition a grid with $\Delta r =
0.000025 \parsec$ covers the shock neighborhood.)
The interior boundary is at~$r = 4.1 \parsec$.
For the $\nh = 50 \percc$ simulation, 
$\Delta r = .001 \parsec$ ($\Delta r/r_s \approx 0.02\%$)
over the bulk of the remnant,
and on the cooling region~$\Delta r$ varies from $0.00013 \parsec$  to
$0.000016 \parsec$ as necessary to keep $\Delta r/L_\cool \approx 1\%$.
(Throughout the shock is resolved by a grid with 
$\Delta r = 0.000004 \parsec$.)
The interior boundary is at~$r = 1.5 \parsec$.

In all cases, the shock velocity changes significantly at the time of shell
formation.  In the case of the rarest medium, after the dense shell forms,
the shock velocity ($v_s \approx 90 \kmps $) is too low
for oscillations to begin.
In the other two cases there are many oscillations before the instability
subsides, near~$v_s \approx 120 \kmps $.  The $\nh = 50 \percc$ simulation
shown in Figure~\ref{case50.0} extends until
we lose sufficient resolution to determine the oscillation
amplitude to within~$10\%$; 
however, we still can continue the evolution to
determine for how long the instability persists.  (The region of the
cooling law with power-law index~$\alpha \approx -2.2$ terminates at a
temperature corresponding to a shock velocity~$v_s = 130 \kmps$.  The
difference between our cutoff value and that of 
Innes et al.~[\shortcite{innes}] 
and of Gaetz et al.~[\shortcite{gaetz-edgar-chevalier}] 
is due to the different
cooling treatment that we adopt.  The slope of our less realistic piecewise
power-law cooling law supports oscillations at lower temperatures.)  The
amplitudes of the oscillations in~$v_s$ are quite large; they lie
in a nonlinear regime in which the amplitudes have reached a saturation
value.
The amplitudes do not appear to depend on the different power laws in
the cooling function.  We note that the oscillations
do not affect the gross behavior of~$r_s$ as a
function of time.

To assess the physical significance of this particular mechanism for
producing fluctuations in the shock velocity, we next consider
several other physical mechanisms for producing shock-velocity
fluctuations.

First we discuss the effect of ``reverberations'' on the shock velocity.  The
birth of the remnant leaves behind waves that propagate back and forth 
inside the remnant.
To investigate the evolution of these waves, we must adopt somewhat
more realistic initial and interior-boundary conditions.  Following Cioffi
et al.~(\shortcite{cmb}), for a revised initial condition
we begin at a time close to the explosion, placing the 
explosion kinetic energy into a ball with constant density (100 times the ISM
density) and fixed total mass ($3\,\msolar$).
We set the velocity distribution to be linear with radius,
$v \propto r$, as might be expected for a collection of particles that have
not yet interacted appreciably with the surrounding cold medium.
The pressure in the ball is set to the ISM value.
(The total thermal energy in the ball is essentially negligible,
over eight orders of magnitude smaller than the total kinetic energy.)
For an interior boundary condition, we choose a ``hard-sphere,'' perfectly
reflecting boundary at the small radius~$r_{\rm inner} = 0.25 \parsec $.

\begin{figure}
\myepsf{6.5in}{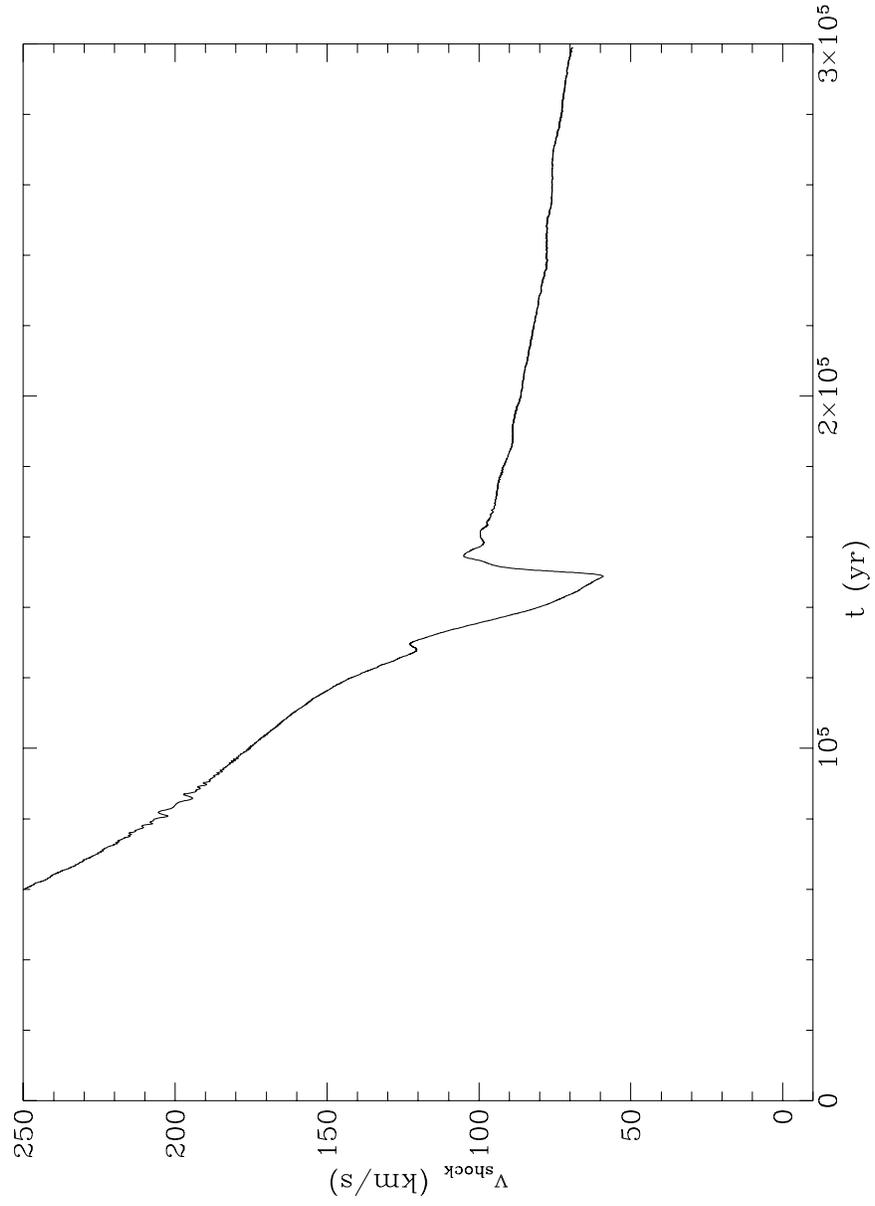}{0}
\caption{Shock velocity for ISM density~$\nh = 0.1$ for calculation
using hard-sphere interior boundary condition.
\label{vs-hardsphere}}
\end{figure}

Figure~\ref{vs-hardsphere}
shows the resulting time evolution of~$v_s$ for the
case of the low-density interstellar medium~($\nh = 0.1 \percc$).
(This simulation uses~$\Delta r = 0.01 \parsec$; in the period shown the
shock radius expands to~$66 \parsec$.)
Well after the mass swept up exceeds the initial explosion mass, only
occasional, small-amplitude fluctuations perturb
the overall evolution of the shock velocity.  
Other deviations from the smooth
evolution of~$v_s$ are similar to those seen in the simulations without the
violent initial conditions 
(in Figure~\ref{case0.1}).

\begin{figure}
\myepsf{6.5in}{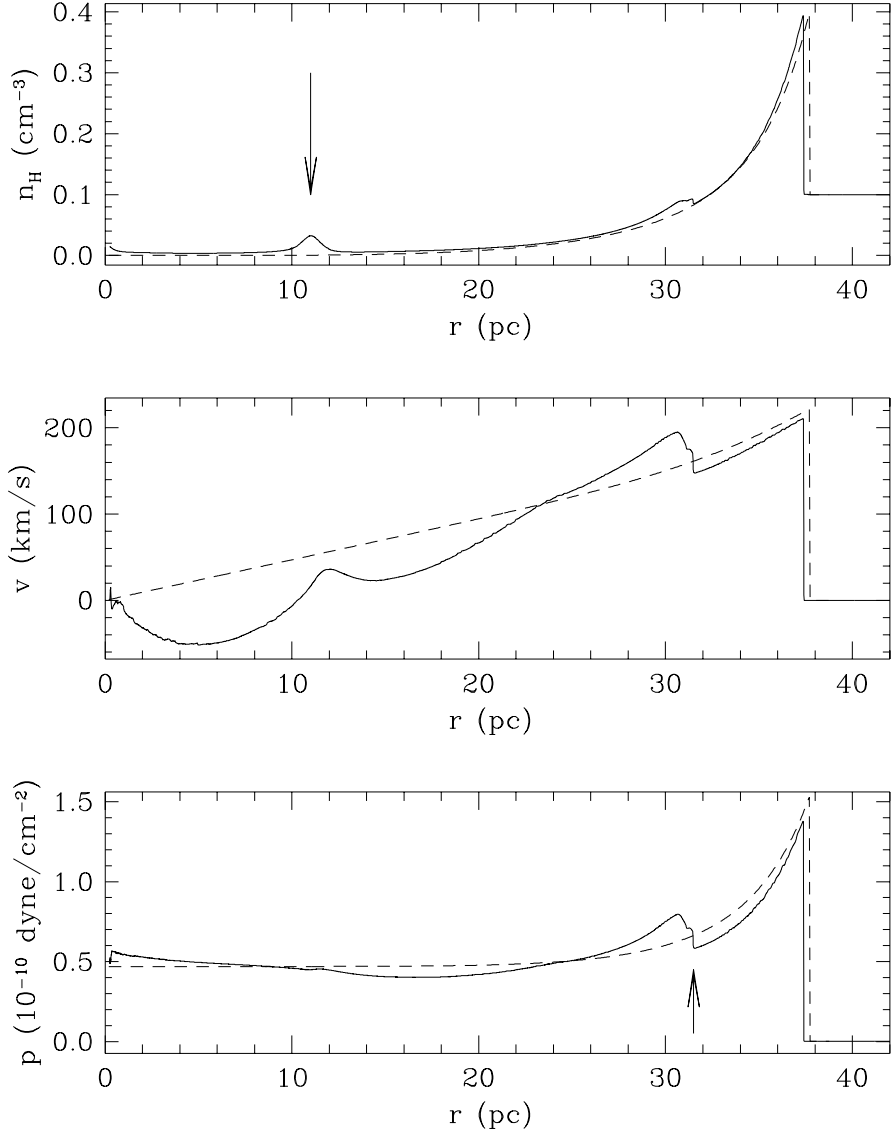}{0}
\caption{Density, velocity, and pressure at~$t = 5 \times 10^4 \yr $
with ISM density~$\nh = 0.1 \percc $
and hard-sphere interior boundary condition.  
The arrow in the top panel points out a density enhancement, a remnant of
the initial conditions; the arrow in the bottom panel points out a
secondary shock.
The Sedov-Taylor similarity
solution at this time is given with dotted lines.
\label{hardsphere}}
\end{figure}

Figure~\ref{hardsphere}
shows the mass density, velocity, and pressure of the flow
at a remnant age of~$5 \times 10^4 \yr $, 
well before the shell-formation time.
In the remnant interior the most apparent differences from the smooth
Sedov-Taylor solution (shown with dotted lines) 
are (1)~inward flow near the center of the remnant,
(2)~a density enhancement (at the arrow in the top panel) 
formed when the shock developed, and 
(3)~a weak shock (at the arrow in the bottom panel) 
traveling toward the main shock.  
The small $v_s$~fluctuation in Figure~\ref{vs-hardsphere}
at~$t \approx 1.2 \times 10^5 \yr $ occurs when the weak shock reaches
the main shock.  Although other waves
in the remnant interior may travel between these features and the main shock
and perturb the progress of the main shock, none have appreciable
amplitude.  We conclude that the effect of reverberations on the evolution
of the shock front must be small, particularly in comparison with the
effects of the radiative instability.
In our simulations such waves may decay in part because of
insufficient numerical resolution, but it is unlikely that they would cause
substantial variation in~$v_s$.  
Some differences from the flows shown in Cioffi et
al.~(\shortcite{cmb}) may arise from differences in initial
conditions.

As a second source of possible shock-velocity fluctuations, we consider
density fluctuations in the upstream gas.  Here we perform
simulations in the simpler plane-parallel geometry.  The cold gas flows into
the computational region from the upstream boundary into the shock and then
into the cooling region.  
This structure ends at a wall at the other boundary.
As discussed at the end of \S\ref{intro}, because our cooling law
has the form~${\cal L} \propto \rho^2 f(T)$, for each inflow velocity 
it suffices to consider just one value of the inflow (interstellar) density,
and we arbitrarily choose~$\nh = 50 \percc $.
For a given mean shock velocity and upstream temperature, 
we can then convert to any value of the upstream density 
by scaling densities and pressures by the
factor~$\nh/50 \percc$ and simultaneously scaling time and lengths by the
factor~$50 \percc/\nh$.

We start with approximately the steady-state solution 
for the given cooling law and allow sinusoidal density fluctuations 
to advect into the shock structure at
two mean inflow velocities (150 and $210 \kmps$).  
Numerically, we simply vary
with time the upstream boundary condition imposed at the fixed computational
boundary.  For simplicity, 
we do not impose numerical flux-limiting techniques
in the upstream flow, where these density fluctuations should merely advect
passively.

We choose the two representative mean velocities (150 and $210 \kmps$) to
represent the range in which we observe the shock instability 
in the spherical calculations.
(Again, these two choices correspond to shock temperatures with different
exponents in our piecewise power-law cooling function.)  
The oscillation period~$\tau$ implies 
a characteristic wavelength~$L_\tau = 4 \left< v_s\right> \tau$ for
incoming perturbations, where~$\left<v_s\right>$ is the mean relative speed
between the shock and the incoming gas.  
(For the $150 \kmps$ case, $L_\tau = 0.1 \parsec$,
and for the $210 \kmps$ case, $L_\tau = 0.8 \parsec$.)
One perturbation of this
length scale passes through the shock front 
in approximately the time taken by
one cycle of the cooling instability.
We choose our fluctuations to have wavelengths~$\lambda$
smaller than and comparable to~$L_\tau$, and with amplitudes
we label ``small''~($\delta n_H/\left<n_H\right> = 1/14$)
and ``large''~($\delta n_H/\left<n_H\right> = 1/2$).
(The effect of upstream density perturbations on certain shock oscillations
was treated in a recent paper by Walder~\&
Folini~[\shortcite{walder-folini}].  They simulated the effect of a single
sinusoidal density lump on oscillations they identify as less violent than
the oscillations we consider here.  The results, however, are
qualitatively consistent with ours.)

\begin{figure}
\myepsf{6.5in}{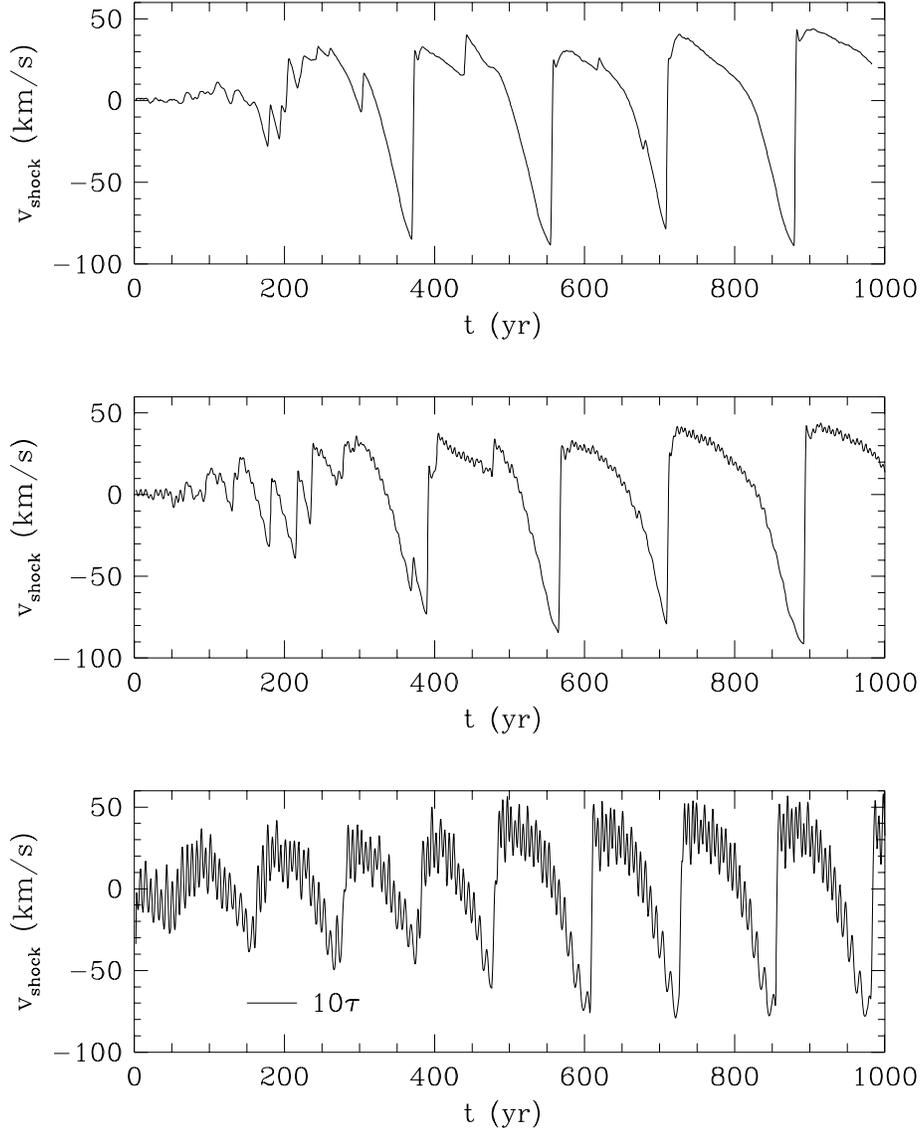}{0}
\caption{Evolution of shock velocity for a shock moving with mean
velocity $150 \kmps$.  
We show the effect of short-wavelength upstream density
perturbations by comparing the cases of (top)~no, (middle)~small, and
(bottom)~large-amplitude fluctuations.  The line segment in the
bottom panel indicates an interval 10 times the fluctuation timescale~$\tau$.
\label{small-lambda-150}}
\end{figure}

\begin{figure}
\myepsf{6.5in}{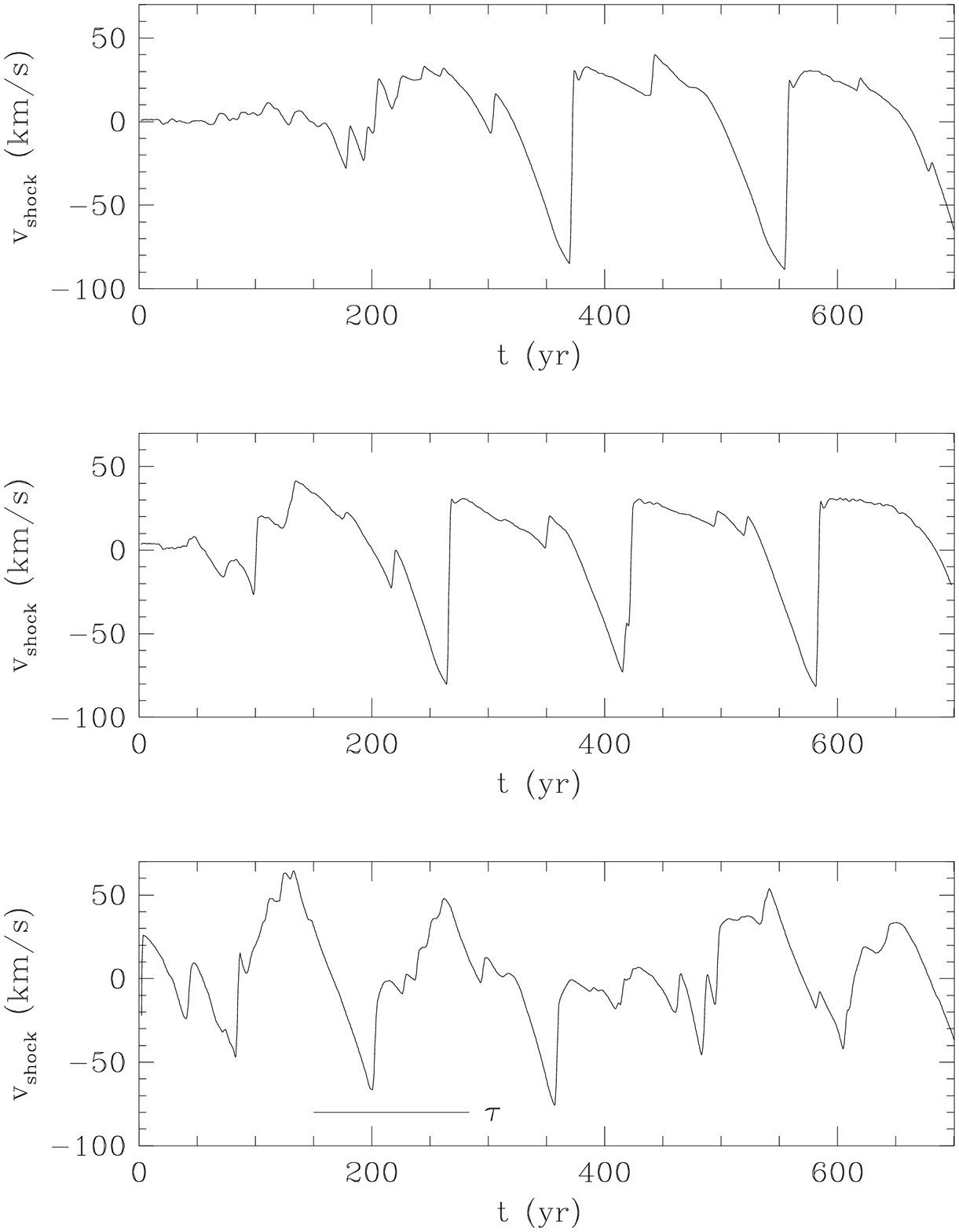}{0}
\caption{Evolution of shock velocity for a shock moving with mean
velocity~$150 \kmps$ perturbed by moderate-wavelength upstream density
fluctuations.  We show the cases of (top)~no, (middle)~small, and
(bottom)~large-amplitude fluctuations.  The line segment in the bottom
panel indicates the fluctuation timescale~$\tau$.
\label{moderate-lambda-150}}
\end{figure}

\begin{figure}
\myepsf{6.5in}{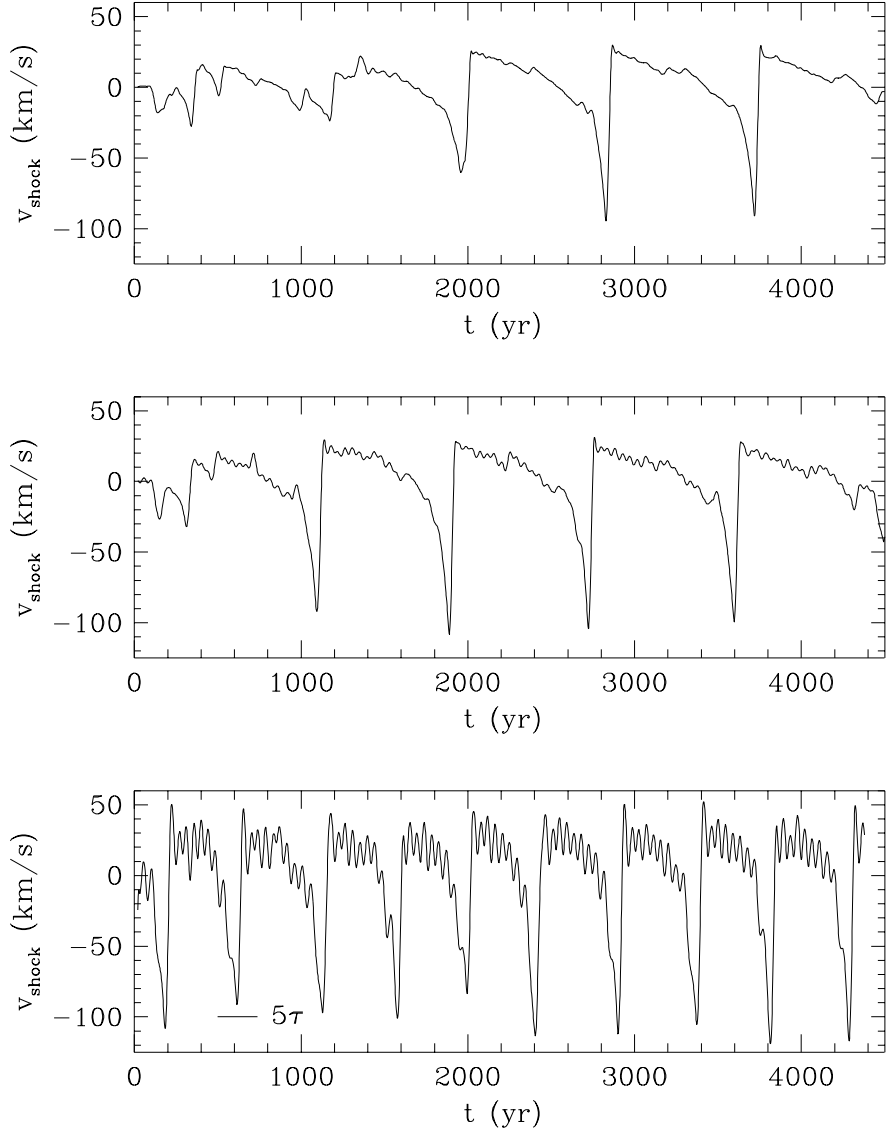}{0}
\caption{Evolution of shock velocity for a shock moving with mean
velocity~$210 \kmps$ perturbed by short-wavelength upstream density
fluctuations.  We show the cases of (top)~no, (middle)~small, and
(bottom)~large-amplitude fluctuations.  The line segment in the bottom
panel indicates a time interval five times the fluctuation timescale~$\tau$.
\label{small-lambda-210}}
\end{figure}

\begin{figure}
\myepsf{6.5in}{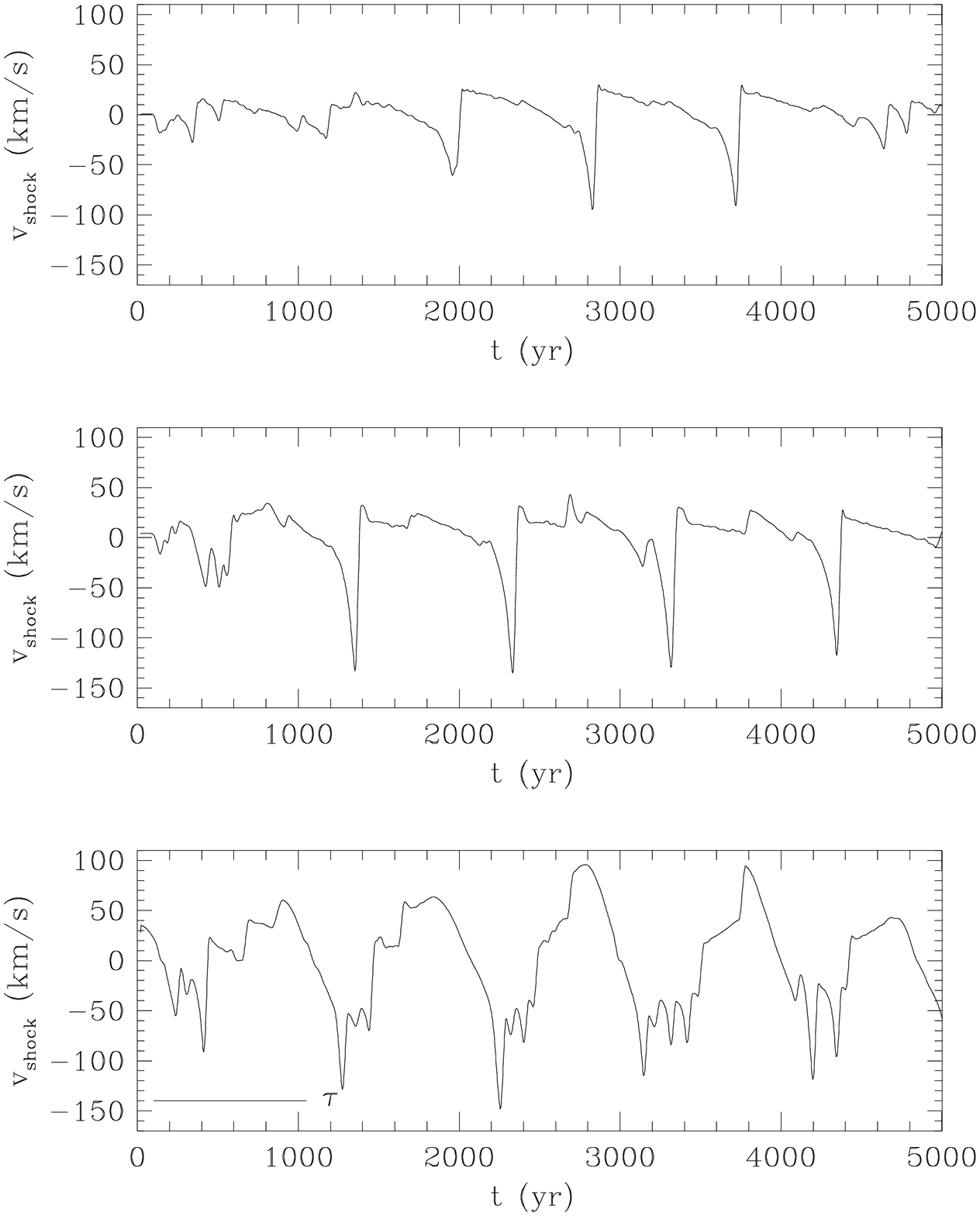}{0}
\caption{Evolution of shock velocity for a shock moving with mean
velocity~$210 \kmps$ perturbed by moderate-wavelength upstream density
fluctuations.  We show the cases of (top)~no, (middle)~small, and
(bottom)~large-amplitude fluctuations.  The line segment in the bottom
panel indicates the fluctuation timescale~$\tau$.
\label{moderate-lambda-210}}
\end{figure}

The next several figures show the resulting behavior of~$v_s$ in the 
cases of no, small, and large-amplitude density
fluctuations for the various choices of~$v_s$ and~$\lambda$.
Figure~\ref{small-lambda-150} shows 
the $150 \kmps$, short-wavelength ($\lambda = 0.001 \parsec$) case;
Figure~\ref{moderate-lambda-150},
the corresponding moderate-wavelength ($\lambda = 0.02 \parsec$) case.
These calculations use grid size $\Delta x = 10^{-6} \parsec$ 
($\Delta x/L_\cool^\steady \approx 0.0012$,
where $L_\cool^\steady$ is the steady-state cooling length).
Figures~\ref{small-lambda-210} and~\ref{moderate-lambda-210} 
show the two $210 \kmps$ cases, 
with short ($0.01 \parsec $)
and moderate ($0.2 \parsec$) wavelengths respectively;
these calculations use~$\Delta x = 10^{-5} \parsec$ 
($\Delta x/L_\cool^\steady \approx 0.0014$).
The line segments in each of these figures indicates the fluctuation
timescale~$\tau = \lambda/\meanvs $, where $\meanvs$ is $150 \kmps$ or $210
\kmps$, as appropriate.
The passage of these density fluctuations through the shock is reflected in
increased, extra variation in shock speed.

Short-wavelength
density fluctuations, of small and large amplitude, pass through the shock
front in a time small compared with the cooling-instability period
and merely
vary the shock velocity over the short times required for one wavelength to
pass through the shock.  The amount of gas that passes through the shock in
each cooling-instability period is not changed by these fluctuations, and so
the gross features of the cooling-instability fluctuations are unchanged.  
In the case of the moderate-wavelength fluctuations, however, 
the amount of gas
passing through the shock in each cooling-instability period varies greatly
from period to period.  This is most pronounced when the density fluctuations
have large amplitude, and we see 
(in Figures~\ref{moderate-lambda-150} and \ref{moderate-lambda-210})
that the evolution of the shock velocity is affected qualitatively.  However,
we note that in no case is the amplitude of the shock-velocity fluctuations
reduced.  In the last case, in which there is some qualitative change, the
oscillations are increased, not reduced.  We conclude that these oscillations
should persist in the presence of any upstream density fluctuations.

\section{Conclusions}

The calculations discussed here show that when the cooling instability
is active, the shock velocities of supernova remnants may fluctuate
considerably.
We see further that reverberations in remnant interiors, consequences of
remnants' complicated births, 
are unlikely to cause shock-velocity fluctuations
of this magnitude.  

Density inhomogeneities in the interstellar medium can cause or modify
shock-velocity fluctuations.  
Of course, they may affect the oscillations due to the cooling instability by
affecting the mean shock velocity.  To have a more direct, large-amplitude
effect on oscillations in~$v_s$, however, 
they must be of quite large amplitude themselves. 
Our simulations show that to
have a substantial qualitative effect on the dynamical cooling oscillations,
they must further have 
length scales comparable to~$L_\tau = 4 \left< v_s \right> \tau$, 
where~$\tau$
is the oscillation period.
(That is, the shock must traverse this length scale a time comparable to the
duration of one oscillation.)

It appears clear that the shocks of many supernova remnants undergo the
cooling instability discussed in this paper.  
The results should be generally correct,
although details may be incorrect because of various assumptions: the
computations are one-dimensional, the cooling treatment is highly
simplified and does not take into account nonequilibrium effects, and
the effects of magnetic fields and of radiative transfer are neglected.

In a separate paper (\cite{kc97b}) we discuss simulations incorporating 
a magnetic field
oriented transverse to the direction of shock propagation
(compression at the shock front tends to align the magnetic field into the
plane of the front).  
These agree with the results of T\'oth \&
Draine (\shortcite{toth}), 
which show that magnetic fields can suppress the
global cooling instability by providing a source of non-thermal pressure in
cold, dense gas.  
For cooling laws such as the one we adopt for
130 to $200 \kmps$ shocks, however, 
very large magnetic fields, corresponding to inverse Alfv\'en Mach
numbers~$M_A^{-1} = v_A/v_s$ 
(where the Alfv\'en speed is~$v_A = B/(4\pi\rho)^{1/2}$) 
in excess of~0.3,
would be required to change the qualitative oscillatory behavior.
(We note parenthetically that the presence of a magnetic field can change
the stability properties at a given shock velocity by changing the shock
temperature; this effect may change the velocity at which oscillations cease.
For example, we find that a $150 \kmps$ shock is stable if the 
upstream Alfv\'en velocity is $10 \kmps$.)
For the cooling law adopted, 
we conclude that the general picture we have described should hold even
with typical magnetic-field strengths.

Raymond et al.~(\shortcite{raymond-wallenstein-balick})
reported some observational evidence 
for the instability in the Vela supernova remnant.
We conclude with a short discussion of considerations for identifying other
remnants that may be subject to this instability.

A first requirement for the presence of the instability is for the remnant to
be in the radiative phase.  Fast radiating remnants may readily be
distinguished by the
presence of strong optical emission lines (the main sources of the cooling).
For example, according to models like those of Cox~\&
Raymond~(\shortcite{cox-raymond}), highly temperature-dependent \ion{O}{3}
lines identify radiating shocks with velocities greater than~$100 \kmps$.
In our results, radiating remnants are oscillatory if they have mean shock
speeds greater than~$120 \kmps$, but
thus far, most remnants with observed \ion{O}{3}~lines have inferred
shock velocities $\la 100 \kmps$ or $> 300 \kmps$, outside our range of
interest.  Hester, Raymond, \& Blair~(\shortcite{hester}) find shocks in the
Cygnus Loop remnant with velocities between $130$ and $180 \kmps$, although
they argue that the cooling has not progressed far enough for instabilities
to have a significant effect.  Of course a precise determination of shock
velocities from line emissivities may be hampered by strong differences
between the predictions for oscillating cooling columns and for steady-state
profiles at the same propagation speed, but nonetheless
the strong optical line emission should still be present (\cite{innes}).

A different method for determining shock velocities was reported by Koo \&
Heiles~(\shortcite{koo-heiles}), who made
\ion{H}{1} 21-cm observations of a collection of
northern remnants.
Although their observations were hampered by confusion with the Galactic
\ion{H}{1}~background, in many remnants
they found high-velocity \ion{H}{1} gas.  This gas is presumably accelerated
by supernova blast waves, and if so it measures the mean remnant expansion
speed.
Several expansion velocities exceed $120 \kmps$
and hence may identify objects with shock waves in our range of interest:
G~117.4+5.0, HB~21, and OA~184.
(The remnant CTA~1 is also listed as having an appreciable expansion
velocity, but more recent observations [\cite{pinneault-et-al}] suggest
that the actual velocity is lower.)

Where expansion velocities are not well known, we may identify some candidate
objects by seeking the remnant sizes that correspond to the
shock velocities of interest.
We note from Figure~\ref{parameters}
that for a given remnant environment (specified by some
value of the ambient density~$\nh$) the range of shock radii for which
oscillations may be observed is rather limited.  If we consider a range of
``typical'' interstellar densities, however, for example,
$0.1 \percc < \nh < 10 \percc $, we find a wide range of shock radii for
which the remnants are candidates for
oscillations:~$10 \parsec \la r_s \la 30 \parsec$ 
(assuming~$E_{51} \approx 1$).
Almost half of the 24 Galactic supernova remnants listed by
Green~(\shortcite{green84}, \shortcite{green88}, \shortcite{green91}) 
as having reasonable
distance estimates may lie in this range, and so may have shock velocities
that subject them to oscillations.  The
Vela supernova
remnant is selected by this simple criterion; the other objects on this
short list are the remnants CTB~37A, CTB~37B, CTB~87, G~320.4$-$01.2,
IC~443, Kes~67, and VRO~42.05.01.  We have excluded several with sizes
in the range of interest: they are believed to be too young (SN~1006
and RCW~86 [possibly $=$ SN~185]), observations indicate
expansion velocities that are too low (CTB~1 [\cite{hailey-craig}]) or too
high (W44 [\cite{koo-heiles-1995}]), or they may still be in the adiabatic
stage (HB~3 [\cite{leahy}]).



\acknowledgements

We thank Edwin Salpeter and the anonymous referee for useful comments on
earlier revisions of this paper.

This research has been carried out at Cornell University with the generous
support of the NSF~(AST-9119475) and NASA~(NAGW-2224)
under the LTSA~program.

Some computations reported herein were carried out using the resources of
the Cornell Theory Center, which receives major funding from the~NSF and
New York State, with additional support from~ARPA, the National Center for
Research Resources at the~NIH, IBM Corporation,
and other members of the Center's Corporate Partnership Program.

\clearpage


\appendix

\section{Numerical Methods}

The equations of motion for an inviscid one-dimensional cooling gas are
\begin{eqnarray}
\label{density-eom}
\pderiv{}{t}[r^d\rho] & = & - \pderiv{}{r} [r^d\rho v], \\
\label{momentum-eom}
\pderiv{}{t}[r^d\rho v] &
              = & -\pderiv{}{r}[r^d\rho v^2] - d r^d\pderiv{p}{r}, \\
\label{energy-eom}
\pderiv{}{t}[r^d e_{total}] & 
              = & -\frac{1}{r^d} \pderiv{}{r} [r^d(e_{total} + p)v]
                    - r^d \nh^2 L(T),
\end{eqnarray}
where~$\nh^2 f(T)$ gives the rate of energy loss per unit volume, and $d = 0$
for the plane-parallel geometry and $d = 2$ for the spherical geometry.
The differencing scheme is based on these equations.  We use operator
splitting
to treat the hydrodynamic and cooling terms separately.

\subsection{Hydrodynamics}

First we describe the treatment of the hydrodynamics.  
The pertinent terms on
the right-hand sides of 
equations~(\ref{density-eom})--(\ref{energy-eom}) fall
into two classes: 
(1)~the flux terms, expressible as pure spatial derivatives,
and (2)~the momentum source term~$r^2 \partial p/\partial r$ in spherical
symmetry.  To calculate the
numerical fluxes corresponding to the source terms, 
we use an implementation of
flux-corrected transport~(FCT)~(\cite{zalesak}).  The FCT method switches
between high and low-order numerical fluxes as required to enforce particular
variation requirements.  For the source term we use simple forward
differencing ($u^{n+1} = u^n + \Delta t g(u^n)$, where~$g(u)$ specifies the
source term as a function of the flow variables).

The FCT flux-calculation method as described by Zalesak~(\shortcite{zalesak})
requires calculating high and low-order numerical fluxes.  
For these we use the
Lax-Wendroff and Lax-Friedrichs schemes~(\cite{lw}; \cite{richtmyer}). 
The method favors the high-order flux (for accuracy) but switches to the
low-order flux where necessary to keep the updated variables within specified
bounds.  Zalesak recommends choosing these bounds in the following way:
Calculate provisional, ``diffused'' updates~$\tilde{u}^{n+1}$ 
using the low-order fluxes.  
Set the bounds for the updated values~$u_j^{n+1}$ by the minimum and
maximum values in the set~$\{ u_{j-1}^n, u_j^n, u_{j+1}^n,
\tilde{u}_{j-1}^{n+1}, \tilde{u}_j^{n+1}, \tilde{u}_{j+1}^{n+1} \}$.  
We adopt this choice, 
except that we remove the geometric factors~$(r_{j \pm 1}/r_j)^d$
from the values at the neighboring gridpoints.

When applied to inviscid fluid flow, the Lax-Wendroff scheme may develop
unphysical features such as oscillations and negative pressures.  At low to
moderate Mach numbers, the FCT method removes this unphysical behavior.
At higher Mach numbers, however, negative pressures can arise near
discontinuities and regions of strong gradients. 
We prevent this by further trying to enforce a lower bound 
on the internal-energy density (proportional to the pressure).

This internal-energy bounding proceeds in the following complicated way:
The updated values of the flow variables are given by
\begin{equation}  \label{antidiffuse}
  u_j = \tilde{u}_j - \frac{1}{\Delta x} (c_{j+1/2}A_{j+1/2} -
                                           c_{j-1/2}A_{j-1/2}),
\end{equation}
where~$\tilde{u}_j$ is the low-order, ``diffused'' provisional value and the
``antidiffusive fluxes''~$A_{j \pm 1/2}$ are defined 
as the differences between
the high- and low-order fluxes.  The correction factors~$c_{j \pm 1/2}$ lie
between~$0$ and~$1$ and are determined by the flux-calculation method.  
(These
correction factors are chosen to be functions of the gridpoint but not of the
particular quantity being calculated:  the same value is chosen for all three
equations.)  For quantities evolved
according to flux and source terms alone, Zalesak gives the algorithm for
calculating the correction factors~$c_{j \pm 1/2}$.

When written in the form of equation~(\ref{antidiffuse}), the updated values
can be regarded as functions of 
the two correction factors:~$u_j = u_j(c_{j - 1/2}, c_{j + 1/2})$. 
The internal-energy density~$\einternal$ is a known function of
the basic flow variables~${u_j}$, 
and so we likewise regard it as a function of the
two correction factors.  

If the partial derivatives~$\partial \einternal/\partial c_{j \pm 1/2}$ have
the same sign, then we find~$\tilde{c}$, the largest allowed value of~$c$
between~$0$ and~$1$ such that~$\einternal(\tilde{c}, \tilde{c})$ is greater
than the imposed minimum value.  Then we test this value of~$\tilde{c}$ by
evaluating~$\einternal (\tilde{c}, 0)$ and~$\einternal (0, \tilde{c})$.  
In the
(unlikely) event that either of these is disallowed (i.e., less than the
minimum acceptable value), we reduce~$\tilde{c}$ so that these two test
quantities lie in the accepted range.

Next, consider the other case, 
in which only one of the two partial derivatives is negative.  
The simulations described in this paper employ the following method:
for specificity, let the negative 
partial derivative be~$\partial \einternal/\partial c_{j-1/2}$.  
We find~$\tilde{c}_{j-1/2}$, the largest value of $c_{j-1/2}$
between~$0$ and~$1$ 
such that~$\einternal (c_{j-1/2}, 0)$ lies in the acceptable range.  
Then we test this value of~$\tilde{c}_{j-1/2}$ by
evaluating~$\einternal(\tilde{c}_{j-1/2}, 1)$.  
In the (unlikely) event that it
falls outside the acceptable range, 
we reduce~$\tilde{c}_{j-1/2}$ so that this
test quantity falls within the acceptable range.

(In other simulations, such as those incorporating a one-dimensional
magnetic field---for example, those described in Kimoto \& Chernoff
[\shortcite{kc97b}]---we find that the above method does not prevent
negative internal energies, and so we use the following alternative method:
We find the two quantities~$\bar{c}_{j \pm 1/2}$, 
the largest values of $c_{j \pm 1/2}$
between $0$ and $1$ such that $\einternal (c_{j-1/2}, 0)$ and $\einternal
(0, c_{j+1/2}$) lie in the acceptable range.  Then we test these values by
evaluating~$\einternal (\bar{c}_{j-1/2}, \bar{c}_{j+1/2})$.  In the
event that it lies outside the acceptable range, we use instead the values
of~$\tilde{c}_{j \pm 1/2}$, determined so that
$\einternal(\tilde{c}_{j-1/2}, \bar{c}_{j+1/2})$ and 
$\einternal(\bar{c}_{j-1/2}, \tilde{c}_{j+1/2})$ are acceptable.)

This prescription calculates values of the correction factors~$c_{j \pm 1/2}$
that, in the vast majority of cases, ensure that the internal-energy
density~$\einternal$ remains above the imposed minimum value.  (In all of the
simulations discussed in this paper, it ensures that the internal-energy
density remains positive.) We use the
smallest, most conservative value of the correction factors, taken from this
prescription and the standard FCT method.

\subsection{Cooling}

As noted in the previous subsection, we use operator splitting to treat
hydrodynamics and cooling separately.  We use the first-order-accurate
semi-implicit Euler method
to perform the update.  
We fit the radiative cooling law of Raymond et al.~(\shortcite{rcs})
by a piecewise set of power laws.  
(These power laws are calculated to fit the
set of values enumerated below.)
To turn the cooling off at low temperatures~$T \la 2 \times 10^4 \kel$, we
multiply the power-law fit by a function
\begin{equation}
  {\cal T}(T)
      = \frac{1}{2}
          \left[ 1+ 
                 \tanh  \left(
                 \frac{T - 1.75 \times 10^4 \kel}{860 \kel} \right)
               \right].
\end{equation}

\begin{tabular}{ll}
$T$ (K)         & cooling rate ($\erg {\rm \, cm}^3 {\rm \, s}^{-1}$) \\
\hline
$1.0 \times 10^4$ & 4.0  \\
$1.5 \times 10^4$ & 22.0 \\
$3.2 \times 10^4$ & 14.0 \\
$1.0 \times 10^5$ & 68.0 \\
$2.5 \times 10^5$ & 15.0 \\
$5.0 \times 10^5$ & 15.0 \\
$2.0 \times 10^6$ & 13.0 \\
\end{tabular}

\subsection{Multigrid}

In order to calculate regions of particular interest accurately
(i.e.,~near the shock and the dense shell), we use a hierarchy of grids with
different grid sizes.  Berger~\& Colella~(\shortcite{berger}, hereafter BC)
describe a method for solving systems of conservation laws on locally refined
grids.  (They describe both the method for working with a hierarchy of grids
and an algorithm to determine where local mesh refinement is required.  We
employ only the former ideas.)

The BC method integrates local high-resolution calculations nested
within larger, lower-resolution calculations.  It maintains the global
conservation laws, as required to ensure the correct evolution of
discontinuities such as shocks.  We borrow their method for maintaining
multiple grids.

Within the BC framework one must often interpolate values from
coarse grids onto fine grids, for example, when introducing new fine grids or
when calculating values at the boundary of existing fine grids.  (The inverse
process, transferring values from fine to coarse grids, must be done via
spatial averaging in order to maintain the global conservation laws).  
We adopt
the interpolation method used within the piecewise-parabolic method~(PPM) of
Colella~\& Woodward~(\shortcite{colella}; \cite{woodward}).  This method
produces interpolating parabolas from gridpoint values.  We enforce a
monotonicity constraint, as described by Colella~\& Woodward, but no more
complicated modifications (such as their discontinuity detection).

The PPM interpolation method ensures 
that the values of conserved quantities on
fine grids are completely consistent with their values on the corresponding
coarse grids.  (That is, the transfer of values back,
from fine to coarse grids, yields the same coarse-grid values.)  
The method, however, occasionally yields
unphysical values for fine-grid quantities.  
Once again, the problem is usually
negative pressures.  In the calculations we made for this paper, such
calculated unphysical values arose only in the region
just outside the boundaries of fine grids.  (The use of values just outside
fine-grid boundaries is the way that 
fine-grid boundary conditions are enforced
in the BC formalism.)  When these unphysical values arise, we replace the
interpolated quantities by physically allowed values, 
obtained by interpolating
the ``primitive'' variables~$\rho$,~$v$, and~$p$ directly.  
(We further require that
the interpolating parabolas for~$\rho$ and~$p$ remain positive.)  
Because these
are not the conservation-law variables, we lose ``consistency'' (in the sense
defined above) between coarse and fine grids.  
This does not affect the treatment of the global conservation laws,
however, because these values are used merely for boundary conditions,
and do not lie within the fine grids proper.

BC give a sophisticated algorithm for determining 
where to place refined grids,
but for our purposes these are not required.  In our simulations the region
most in need of high resolution is always the cooling region just behind the
main shock.  Within this region, most important are its two edges, namely 
(1)~the shock
itself and (2)~the region of rapid cooling and strong density gradient.  
It is
computationally simple to determine the location of both these features by
searching inward from the upstream boundary.  (We locate the shock by its
characteristic pressure jump, 
and we locate the steep density gradient by some
large value of the mass density that it must encompass.  Of course,
the latter criterion does not apply before the dense shell forms.)
The subgrid for each refined region has a specified size,
and on each time step a grid's subgrids may be translated to follow the motion
of the appropriate features.

Refinement algorithms (such as presented by BC) may determine how much
refinement should be applied, but we do not use these methods, either.  
For our purposes it suffices to choose a fixed grid size 
for each refined level within
a computation.  It is still possible for the program user to change the
refinement of levels by pausing the evolution program and modifying the
representation of the flow,
but naturally it is practical
to do this only at isolated times.  In the spherically symmetric computations
the mean cooling length decreases steadily as the shock velocity 
(and hence the
shock temperature) decreases; as it shrinks such intervention to decrease the
grid size can be necessary in order to maintain sufficient resolution on the
region. 

\subsection{Time step}

For each time step of the coarsest grid in the grid hierarchy (which
corresponds to multiple time steps for its subgrids), we calculate
a new value of the time step~$\Delta t$.
For hydrodynamic stability we obey the
Courant-Friedrichs-Lewy  condition (\cite{richtmyer}) by calculating
\begin{equation}
  (\Delta t)_{\rm CFL}
          = 0.9 \min_{\rm grid} \frac{\Delta r}{\left|v\right| + c}
\end{equation}
on the coarsest grid {\em only.}
For accuracy of the cooling treatment, we calculate
\begin{equation}
  (\Delta t)_\cool
           = 0.1 \min_{\rm grid} \frac{\einternal}{\nh^2 L(T)},
\end{equation}
on all parts of
{\em each\/} grid that is not covered by another, more highly refined grid.
The time step used, then, 
is~$\Delta t = \min [(\Delta t)_{\rm CFL}, (\Delta t)_\cool ]$.

\subsection{Boundary conditions}

We enforce several types of boundary conditions.  All are straightforward to
implement numerically.  Our computational framework requires specification of
input fluxes entering the edge zones so that the recipe for their updates is
the same as that for interior zones.  The enforcement of boundary conditions,
then, requires methods for determining fluxes entering the
computational region from the outside.

First we describe the upstream (ISM) boundary condition.  In this region we
know exactly what values the flow variables should take at positions off the
computational region.  
Thus we can define several (phantom) zones off the edge
of the computational region.  The edge fluxes can then be computed from these
and the ``real'' zones inside the computational region, just as the fluxes in
the interior are computed.

Next we describe the Sedov-Taylor interior boundary condition.  
Here we are given particular values to impose at a particular position.
In this case
we simply define the flux entering that zone 
from the (phantom) exterior of the
computational region to be whatever value is required
so that the updated values in that zone take on the desired values.

Last we describe the perfectly reflecting boundary condition.  
We impose this condition at the position of the gridpoint itself.  
As in the previous situation, we choose the (phantom) momentum flux 
entering the edge zone such that the velocity there vanishes.  
For the other phantom fluxes, we set them equal in magnitude 
but opposite in sign to the fluxes entering the zone 
from the interior of the computational region,
since that is equivalent to having the zone divided in half by a wall (which
of course allows no fluxes through it).

\clearpage



\end{document}